\documentclass{ws-procs9x6}
\usepackage{graphicx}
\newcommand\beq{\begin{equation}}
\newcommand\eeq{\end{equation}}
\newcommand\beqa{\begin{eqnarray}}
\newcommand\eeqa{\end{eqnarray}}
\begin{document}

\title{Screening effect in Quark-Hadron \\
Mixed Phase}

\author{T. TATSUMI}

\address{Department of Physics, Kyoto University, \\ 
Kyoto 606-8502, Japan\\ 
E-mail: tatsumi@ruby.scphys.kyoto-u.ac.jp}

\author{D.~N. VOSKRESENSKY}

\address{Moscow Institute for Physics and Engineering, \\ 
Kashirskoe sh. 31, Moscow 115409, Russia}  


\maketitle

\abstracts{Possibility of the structured mixed phases at first order
phase transitions in neutron stars is reexamined by taking into 
account the charge
screening effect. The Maxwell construction is shown to be not 
conflicted
with the Gibbs conditions once the Coulomb potential is properly taken
into account.
Taking the hadron-quark deconfinement transition as an
example, we explicitly demonstrate a mechanical instability of 
the geometrical
structure of the structured mixed phase by the charge screening
effect. 
In this case we have effectively the 
picture 
given by the
Maxwell construction.}

\section{Introduction}

It is now commonly accepted that various phase transitions may occur in 
compact star interiors or during the gravitational collapse from
progenitor stars. Possibilities of the meson (pion and kaon)
condensations 
and the hadron-quark deconfinement transition at high-density matter or
the liquid - gas transition at subnuclear density 
have been studied by many authors. These phase transitions may have some
implications to compact star phenomena, and it has been expected that 
recent progress in observations might reveal such new forms of matter.

Such phase transitions are of the first order in most cases and 
the Maxwell construction has been applied to get the equation of state
(EOS) in phase equilibrium; there appears a separation of 
spatially bulk phases in the mixed-phase with the equal
pressure. Glendenning 
demonstrated a 
possibility of the structured mixed phases (SMP) in such systems 
by invoking the proper treatment based on the Gibbs conditions
\cite{gle}, where the charge density as well as the baryon-number density
are inhomogeneous. 
Subsequently, many authors have demonstrated 
energetic preference of SMP and its existence in a wide density region,
disregarding effects of inhomogeneity of the particle 
configurations and/or the electric field \cite{smp}.  
The geometrical structure of SMP
looks like droplets, rods or slabs as in the nuclear pasta phase 
\cite{pet,nor}.

The Gibbs conditions require the pressure balance and the equality of
the chemical potentials between two phases, denoted by $I$ and $II$, 
for phase equilibrium \cite{gug}.
\footnote{We consider here matter at zero temperature}
 For a multi-component system with more than
one chemical potential, as is common in neutron-star matter, we must
impose the equality condition for each chemical potential in order 
to fulfill  the condition of the 
physico-chemical equilibrium. More definitely, we,
hereafter, consider the charge chemical potential ($\mu_Q$) and 
the baryon-number
chemical potential ($\mu_B$) respecting two conservation laws in neutron-star
matter: $\mu_Q^I=\mu_Q^{II}$ and $\mu_B^I=\mu_B^{II}$. On the other
hand, the first condition is not fulfilled in the Maxwell construction,
since the $\it local$ charge neutrality is implicitly imposed, while
only the {\it global} charge neutrality must be satisfied. When we
naively apply the Gibbs conditions instead of the Maxwell
construction, we can see that there appears SMP in a wide density region and
there is no constant-pressure region in EOS.

SMP, if exists, may have phenomenological implications on compact stars
through e.g., glitches, neutrino opacity, gamma-ray burst or mass of
hybrid stars. 

In this talk we address a controversial issue about the relevance of SMP, 
by taking 
the hadron-quark deconfinement transition as an example \cite{vos}. 
We shall see
that the Debye screening effects greatly modify the 
mechanical stability of SMP.
In the absence 
of SMP we effectively recover the picture of phase equilibrium 
given by the Maxwell
construction where two bulk phases are separated without 
spoiling the Gibbs conditions.

\section{Bulk calculations and finite-size effects}

\begin{figure}[h]
\begin{center}
\includegraphics[width=5cm,clip]{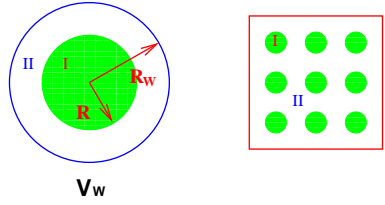}
\end{center}
 \caption{Equal droplets of the phase I embedded in the phase II (right
 panel), and the geometrical structure of the Wigner-Seitz cell (left panel)}
\label{ws}
\end{figure}

Consider SMP consisting of two phases I and
II, where we assume spherical droplets of phase I with the radius $R$ to be
embedded in the matter of phase II and two phases are clearly separated
by sharp boundaries. We divide the whole space into the
equivalent Wigner-Seitz cells with the radius $R_W$ (see Fig.\ref{ws}).
The volume of the cell is $V_W=4\pi R_W^3/3$ and that of the
droplet is $V=4\pi R^3/3$. 

\begin{figure}[h]
\begin{center}
\includegraphics[width=5cm,clip]{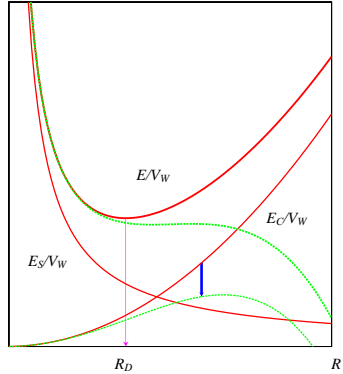}
\end{center}
\caption{Schematic view of the total energy and each contribution in the bulk calculations
 (solid curves). Screening effect reduces the Coulomb energy, shown by
 the thick arrow.}
\label{bulk}
\end{figure}

A bulk calculation proceeds as follows \cite{pet}.
For a given volume fraction factor
$f=(R/R_W)^3$, the total energy $E$ may be written as the sum of the
volume energy $E_V$, the Coulomb energy $E_C$ and the surface energy $E_S$,
\beq
E=E_V+E_C+E_S.
\label{energy}
\eeq
We further assume, for simplicity, that baryon number ($\rho_B^\alpha$) and
charge ($\rho_Q^\alpha$) densities are uniform in each phase
$\alpha,~\alpha=I,II$. Then, $E_V$ can be written as
$E_V/V_W=f\epsilon^I(\rho_B^I)+(1-f)\epsilon^{II}(\rho_B^{II})$ in terms
of the energy densities $\epsilon^\alpha,~\alpha=I,II$. The surface energy $E_S$ may 
be represented as $E_S/V_W=f\times 4\pi\sigma/R$ in terms of the surface
tension $\sigma$. The Coulomb energy $E_C$ is given by 
\beq
E_C/V_W=f\times \frac{16\pi^2}{15}\left(\rho_Q^I-\rho_Q^{II}\right)^2R^2.
\eeq
The optimal value of $R_D$ is determined by the minimum condition,
\beq
\left.\frac{\partial (E/V_W)}{\partial R}\right|_{f}=0,
\eeq
for a given $f$ (see Fig. \ref{bulk}). Since $E_V$ does not depend on $R$, 
we can {\it always} find a minimum as a result of the competition
between the Coulomb and the surface energies, satisfying the 
well-known relation, $E_S=2E_C$.

However, such bulk calculations have been proved to be too crude for the 
discussions of SMP. Instead, a careful consideration of 
the interface of two phases is required. 
As a defect of the bulk calculations they ignore the {\it finite size
effects}. In particular, they have the inconsistent treatment
of the Coulomb potential; they do not use the Poisson equation, so that
the charge density profiles are assumed ab initio to be constants
and the Coulomb
potential is assumed to be $1/r$. If one properly solves the Poisson
equation, one should have the screening effect as a result of the rearrangement
of the charge-density distribution. Hence,  
the radius $R_D$ should be not too large, compared with the Debye
screening length $\lambda_D^{-2}=\sum_i(\lambda_D^i)^{-2}$, 
\beq
1/\lambda_D^{i2}=4\pi Q_i\frac{\partial\rho_{ch}}{\partial\mu_i},
\eeq 
in order the above treatment to be justified,
the suffix $i$ runs over the particle species. 
Otherwise, the Coulomb
energy is reduced by the screening effect, 
which should lead to a {\it mechanical
instability} of SMP in some cases (Fig.~\ref{bulk}).
In the case of the hadron-quark deconfinement transition,
$
\lambda_D^q\simeq 5{\rm fm}
$
and $\lambda_D^p, \lambda_D^e$ are of the same order as $\lambda_D^q$, for
a typical density with $\mu_B\simeq 1$GeV. We
shall see in the following that $R_D$ is typically of the same order as
$\lambda_D\sim \lambda_D^q$, and the mechanical stability of the droplet
is much affected by the screening effect.

\section{Mechanical instability of the geometrical structure of SMP}

\subsection{Thermodynamic potential for hadron - quark deconfinement 
transition}

In the following we consider thermodynamics for non-uniform systems. The
situation is the same as described in Fig.~\ref{ws}: the phase I in the
domain $D^I$ 
consists of $u,d,s$ quarks and electrons and the phase II in the domain
$D^{II}$ neutrons,
protons and electrons. These phases should be clearly separated by the
narrow boundary layer $D_S$ with the width $\sim d_s\leq 1$ fm 
due to the
non-perturbative effect of QCD. We treat such narrow boundary as the
sharp one ($\partial D$) with the surface tension parameter 
$\sigma_{\rm QCD}$ by using
the bag model picture, while the value of $\sigma_{\rm QCD}$ is 
poorly known. We
shall see that the Debye screening length $\lambda_D$ is much 
longer than $d_s$ and
thereby the introduction of the sharp boundary should be reasonable
\footnote{This treatment is also similar to the {\it Gibbs geometrical
surface} \cite{gug}.}. 

Then, the thermodynamic potential per cell
is given by a density functional \cite{par},
\begin{eqnarray}
\Omega=E[\rho ] - \mu_i^{\rm I}\int_{D^{\rm I}}
d\vec{r}\rho_i^{\rm I}- \mu_i^{\rm II}\int_{D^{\rm II}}d\vec{r}
\rho_i^{\rm II},
\label{omeg}
\end{eqnarray}
where $E[\rho ]$ is the energy of the cell and consists of four
contributions:
\beq
E
[\rho ] =\int_{D^{\rm I}} d\vec{r} \epsilon^{\rm I}_{\rm kin+str}
[\rho_i^{\rm I}] + \int_{D^{\rm II}} d\vec{r} \epsilon^{\rm
II}_{\rm kin+str} [\rho_i^{\rm II}]+4\pi R^2\sigma_{\rm QCD}+E_V.
\eeq
The first two terms are given by the kinetic and strong interaction
energies, and the Coulomb interaction energy 
$E_V$ is expressed in terms of particle
densities,
\begin{equation}\label{enden2}
E_V=\frac{1}{2}\int d\vec{r}\,d\vec{r}^{\,\prime}
\frac{Q_i \rho_i (\vec{r}) Q_j \rho_j (\vec{r}^{\,\prime}) }{\mid
\vec{r}-\vec{r}^{\,\prime}\mid},
\end{equation}
with $Q_i$ being the particle charge ($Q =-e <0$ for the
electron).

The equations of motion are given by
$\delta\Omega/\delta\rho_i^\alpha=0$ and written as 
\begin{equation}\label{eom}
\mu_i^\alpha =\frac{\partial\epsilon_{\rm kin+str}^\alpha}
{\partial\rho_i^\alpha}- N^{{\rm{ch}},\alpha }_i V^\alpha
(\vec{r})  ,~~~ N^{{\rm{ch}},\alpha}_i =Q_i^\alpha /e,
\end{equation}
with the electric potential  $V^\alpha (\vec r )$: 
\begin{eqnarray}\label{other1}
V(\vec r ) =- \int d\vec{r}^{\,\prime}\, \frac{e Q_i \rho_i
(\vec{r}^{\,\prime}) }{\mid \vec{r}-\vec{r}^{\,\,\prime}\mid }
\equiv\left\{
\begin{array}{ll}
V^{\rm I}(\vec{r}), & \vec{r}\in D^{\rm I}\\ V^{\rm II}(\vec{r}),
& \vec{r}\in D^{\rm II}
\end{array}
\right. .
\end{eqnarray}
Thus chemical potentials $\mu_i^\alpha$ for charged particles have values depending on the
electric state of the phase as well as on its chemical
composition. Actually it is sometimes called the {\it electro-chemical potential}
to stress this fact \cite{gug}.

\subsection{Gauge invariance}

The thermodynamic potential enjoys the invariance under a gauge
transformation, $V(\vec{r})\rightarrow V(\vec{r}) -V^0$ and 
$\mu_i^\alpha\rightarrow \mu_i^\alpha+N_i^{{\rm ch},\alpha}V^0$, with an
arbitrary constant $V^0$. Hence the chemical potential
$\mu_i^\alpha$ acquires physical meaning only {\it after gauge fixing}
\footnote{Note that $V=0$ is a conventional choice in the usual
treatment of uniform matter, while any constant is possible there.}
.

\begin{figure}[h]
\begin{center}
\includegraphics[width=4cm,clip]{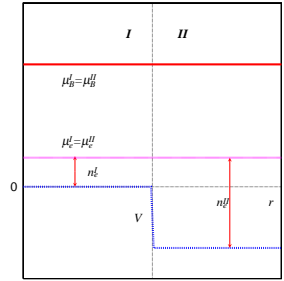}
\end{center}
\caption{Relation between the charge chemical potential $\mu_Q(=\mu_e)$
 and the electron number density $n_e$ in the presence of the Coulomb
 potential $V$. Fulfilling the Gibbs conditions, $\mu_B^I=\mu_B^{II},~\mu_e^I=\mu_e^{II}$, we can change $n_e$ in
 two phases as in the Maxwell construction, if $V$ changes from one
 phase to another.}
\label{gainv}
\end{figure}

Here we reconsider the Gibbs conditions and the Maxwell construction. As
has been  mentioned, on the first glance the Maxwell construction looks 
as contradicting the
Gibbs conditions, especially the equilibrium condition for the charge
chemical potential $\mu_Q(=\mu_e)$ in our context. However, correctly speaking,
when we say $\mu_e^I\neq \mu_e^{II}$ within the Maxwell construction, it means nothing
but the difference in the electron number density $n_e$ in  two
phases, $n_e^I\neq n_e^{II}$; this is because $n_e=\mu_e^3/(3\pi^2)$, 
if the Coulomb
potential is {\it absent}. Once the Coulomb potential is taken
into account, 
using eq.~(\ref{eom}), $n_e$ can be written as 
\beq
n_e^\alpha=\frac{(\mu_e^\alpha-V^\alpha)^3}{3\pi^2}.
\eeq
Thus we may have $\mu_e^I=\mu^{II}_e$ and $n_e^I\neq n_e^{II}$
simultaneously, with the {\it different values of $V$}, $V^I\neq V^{II}$
(see Fig.~\ref{gainv}).

Applying Laplacian ($\Delta$) to the l.h.s. of eq.~(\ref{other1})
we recover the Poisson equation ($\vec{r}\in D^\alpha$),
\begin{eqnarray}\label{qV}
\Delta V^\alpha (\vec{r}) =4\pi e^2\rho^{{\rm
ch},\alpha}(\vec{r})\equiv 4\pi e Q_i^\alpha \rho_i^\alpha
(\vec{r}) . \,\,\,
\end{eqnarray}
The charge density $\rho^{{\rm ch},\alpha}(\vec{r})$  as a
function of $V^\alpha (\vec{r})$ is determined by the equations of
motion (\ref{eom}). Thus eq.~(\ref{qV}) is a nonlinear
differential equation for $V^\alpha (\vec{r})$. The boundary
conditions are
\begin{equation}\label{boun}
V^{\rm I}=V^{\rm II} ,~~~\nabla V^{\rm I}=\nabla V^{\rm II},~~~ {\vec r}\in \partial D\, ,
\end{equation}
where we have neglected a
small contribution of  the surface charge accumulated at the interface of the phases.
We also impose the condition, $\nabla V^{\rm II}=0$, at the boundary of
the Wigner-Seitz cell, which implies that each cell must be charge neutral.

\subsection{Results}

The Debye screening parameter is introduced by
the Poisson equation, if one expands the charge density in $\delta
V^\alpha (\vec{r})=V^\alpha (\vec{r})- V^{\alpha}_{\rm ref}$
around a reference value $V^{\alpha}_{\rm ref}$. Then eq.~(\ref{qV}) renders
\begin{equation}\label{Pois-lin}
\Delta \delta V^\alpha (\vec{r}) =  4\pi e^2
\rho^{{\rm ch},\alpha} (V^{\alpha}(\vec{r})=V^{\alpha}_{\rm ref}
)+ (\kappa^{\alpha} (V^{\alpha}(\vec{r})=V^{\alpha}_{\rm ref} ))^2
\delta V^{\alpha}(\vec{r})+...,
\end{equation}
with the {\it Debye screening} parameter,
\begin{equation}\label{debye}
(\kappa^{\alpha}(V^{\alpha}(\vec{r})=V^{\alpha}_{\rm ref} ))^2 =
4\pi e^2\left[ \frac{\partial\rho^{{\rm ch},\alpha}}{\partial
V}\right]_ {V^{\alpha}(\vec{r})=V^{\alpha}_{\rm ref}} =4\left.\pi
Q_i^{\alpha}Q_j^{\alpha}\frac{\partial\rho_j^{\alpha}}{\partial\mu_i^\alpha}
\right|_{V^{\alpha}(\vec{r})=V^{\alpha}_{\rm ref}}.
\end{equation}
Then we calculate contribution to
the thermodynamic potential of the cell
up to $O(\delta V^{\alpha}(\vec{r}))^{2}$.
The ``electric field energy'' of the cell (\ref{enden2}) can be written 
by way of the Poisson equation (\ref{Pois-lin}) as
\begin{eqnarray}\label{eV}
E_V
=\int_{D^{\rm I}} d{\vec r}\epsilon_{V}^{\rm I} +\int_{D^{\rm
II}}d{\vec r}\epsilon_{V}^{\rm II} =\int_{D^{\rm I}}\frac{(\nabla
V^{\rm I}(\vec{r}))^2}{8\pi e^2}d{\vec r} + \int_{D^{\rm
II}}\frac{(\nabla V^{\rm II}(\vec{r}))^2}{8\pi e^2}d{\vec r},
\end{eqnarray}
that is, in the case of  unscreened approximations, usually called
the Coulomb energy.
Besides the terms given by (\ref{eV}), there
are another contributions arising from effects associated with the
inhomogeneity of the electric potential profile, through implicit
dependence of the particle densities on
$V^{\rm I,II}(\vec{r})$. We will call them ``correlation terms'',
$\omega_{\rm cor}^{\alpha}=\epsilon_{\rm kin+str}^{\alpha}-\mu_i^{\alpha}
\rho_i^{\alpha}.$

We obtain the corresponding
correlation contribution to the thermodynamic potential
$\Omega_{\rm cor} =\int_{D^{\rm I}} d{\vec r}\omega_{\rm cor}^{\rm
I} +\int_{D^{\rm II}} d{\vec r}\omega_{\rm cor}^{\rm II}$:
\begin{eqnarray}\label{om-cor0}
&&\omega_{\rm cor}^{\alpha}= \epsilon_{\rm
kin+str}^{\alpha}(\rho_i^{\alpha}(V^{\alpha}_{\rm
ref}))-\mu_i^\alpha \rho_i^{\alpha} (V^{\alpha}_{\rm ref})-
\rho^{{\rm ch},\alpha} (V^{\alpha}_{\rm ref} )V^{\alpha}_{\rm ref}
\nonumber \\ &&+\frac{V^{\alpha}_{\rm ref}\Delta
V^{\alpha}(\vec{r})}{4\pi e^2} +\frac{
(\kappa^{\alpha}(V^{\alpha}_{\rm ref }) )^2 (\delta
V^{\alpha}(\vec{r}))^2 }{8\pi e^2 }+...,
\end{eqnarray}
where we also used eqs. (\ref{Pois-lin}) and (\ref{debye}). In
general $V^{\rm I}_{\rm ref }\neq V^{\rm II}_{\rm ref }$ and they
may depend on the droplet size.  Their proper choice should
provide appropriate convergence of the above expansion in $\delta
V (\vec{r})$. Taking $V^{\rm I}_{\rm ref }= V^{\rm II}_{\rm ref
}=V_{\rm ref } = const$
we find
\begin{eqnarray}\label{om-cor}
\omega_{\rm cor}^{\alpha}=\frac{ (\kappa^{\alpha}(V_{\rm ref })
)^2 ( V^{\alpha}(\vec{r}) - V_{\rm ref })^2 }{8\pi e^2 },
\end{eqnarray}
except an irrelevant constant.

\begin{figure}[h]
\begin{minipage}{0.55\linewidth}
\begin{center}
\includegraphics[width=6cm,clip]{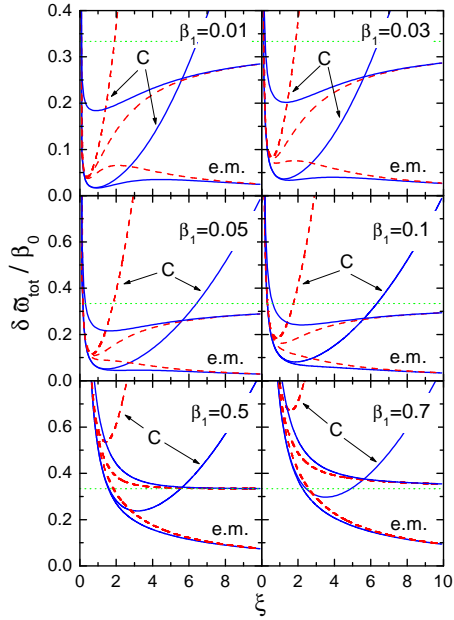}
\caption{Dimensionless value of the thermodynamic potential per droplet
 volume. Solid
lines are given for $f=0.5$ and dashed lines for $f=1/100$. The ratio of
the screening lengths of two phases,
 $\alpha_0=\lambda_D^I/\lambda_D^{II}$, is fixed as one. $\xi$ is a
 dimensionless radius of the droplet, $\xi\equiv R/\lambda_D^I$, with
 $\lambda_D^I\simeq 5$ fm in this calculation. See text for further details. }
\end{center}
\label{effe}
\end{minipage}
\hfill
\begin{minipage}{0.4\linewidth}

~~~~~For given baryon-number \\
chemical potential $\mu_B$ and charge chemical
potential $\mu_Q$, all the particle chemical potentials $\mu_i$ can be
represented in terms of $\mu_B$ and $\mu_Q(=\mu_e)$ with the help 
of the chemical equilibrium
conditions:
\begin{eqnarray}\label{q-c}
\mu_u &-&\mu_s +\mu_e =0, \,\,\,\, \mu_d =\mu_s,\nonumber\\
\mu_n &=&\mu_p +\mu_e ,
\end{eqnarray}
in each phase and 
\begin{eqnarray}\label{q-h1}
\mu_B\equiv \mu_n =2\mu_d +\mu_u ,
\end{eqnarray}
at the boundary. 

Then particle number densities $\rho_i$ are represented
as functions of $\mu_B$, $\mu_Q$ and the Coulomb potential $V$, due to
the equations of motion. Substituting them in the Poisson equation
(\ref{qV}), we can solve it with the proper boundary conditions (\ref{boun}); 
note that $\nabla V=0$ at the boundary of the Wigner-Seitz cell provides
us with another relation between $\mu_B$ and $\mu_Q$.
\end{minipage}
\end{figure}

Thus 
we eventually have the density profiles of all the particles 
for given density or the
baryon-number chemical potential $\mu_B$. 
In Fig.~4 we demonstrate the radius ($R$) dependence of the total thermodynamic
potential  per droplet volume for the case of 
spherical droplets,$\delta \widetilde{\omega}_{\rm
tot}/\beta_0=(\widetilde{\epsilon}_V^{\rm I}+
\widetilde{\epsilon}_V^{\rm II}+ \widetilde{\omega}_{\rm
cor}^{\rm I}+ \widetilde{\omega}_{\rm cor}^{\rm
II}+\widetilde{\epsilon}_{\rm S} )/\beta_0$,   given by the sum of
partial contributions, where tilde denotes each quantity scaled by
the droplet volume $V=4\pi R^3/3$ and $\beta_0$ is a typical quantity with
the dimension of the energy density\cite{vos}. Preparing some wide range for the
value of the surface tension parameter $\sigma_{QCD}$, $\beta_1\propto
\sigma_{QCD}$\cite{vos}, we present two cases of $f$, $f=0.01,0.5$.

The label ``C'' is given for reference to show the previous
non-selfconsistent case, where the Coulomb potential is not screened see
Fig.~\ref{bulk}. We can see that only  
in the limit of $f\ll 1$ and $R\ll \lambda_D^I$, we are able to
recover this case. The ``e.m.'' curve shows the partial contribution to
the thermodynamic potential, 
$\widetilde{\epsilon}_{\rm e.m.}/\beta_0\equiv(\widetilde{\epsilon}_V +
\widetilde{\epsilon}_{\rm S})/\beta_0$, ignoring correlation
terms. Comparing these curves we can see how the screening effect
changes the thermodynamic potential: we can see that the minima at the
``e.m.'' curves disappear already at $\beta_1>0.03$, 
corresponding to unphysically
small $\sigma_{QCD}\sim$ several MeV$\cdot$fm$^{-2}$. 
However, the correlation energy
gives a sizable contribution to allow the minimum for larger value of
$\sigma_{QCD}$. Consequently, the minimum totally disappears between
$\beta_1=0.1$ and $\beta_1=0.5$, which may be interpreted as 
$10<\sigma_{QCD}<50$(MeV$\cdot$fm$^{-2}$) in this calculation. Thus we
have seen a {\it mechanical instability} of the droplet for the medium values
of $\sigma_{QCD}$, which might be in the physically meaningfull range. 

\section{Summary and Concluding remarks}

In this talk we addressed an issue about SMP at the first order phase
transitions in multicomponent systems, like in neutron-star matter. 
We have studied a so called
``contradiction'' between the Gibbs conditions and the Maxwell construction
extensively discussed in previous works. We have 
demonstrated that this 
contradiction is resolved
if one correctly takes into account the difference in 
the ``meaning'' of the chemical potentials used in the two
approaches: the different values of the electron
chemical potentials in the Maxwell construction and the ones
used in the Gibbs conditions do not contradict each other if one
properly
takes into account the electric field.

We have presented a framework based on the density functional theory to
describe thermodynamics in the non-uniform systems. The Coulomb
potential is properly included and particle density profiles are
consistently determined with the Poisson equation. 

Taking the hadron-quark deconfinement transition in high-density matter
as an example, we have demonstrated the importance of the Debye
screening effect, which is a consequence of the above treatment. With a
numerical example, we have seen that the screening effect gives rise to
a mechanical instability for realistic values of the surface tension
parameter of $\sigma_{QCD}$. In this case we may effectively recover the
picture 
given by the Maxwell construction, where the
{\it phase separation} of two bulk phases arises.

Our framework is rather general and it may be applicable to any first 
order phase transition, e.g. the liquid-gas phase transition at
subnuclear density \cite{maru}.






\section*{Acknowledgments}

We acknowledge with special thanks contributions of our collaborators, 
E.E. Kolomeitsev, T. Maruyama, S. Chiba, T. Tanigawa
and T. Endo to the results presented in this paper. The present work of T.T. 
is partially supported by the 
Japanese Grant-in-Aid for Scientific Research Fund of the Ministry
of Education, Culture, Sports, Science and Technology (11640272,13640282)


\begin{thebibliography}{0}
\bibitem{gle} N.K. Glendenning, Phys. Rev. {\bf D46} (1992) 1274; 
Phys. Rep. {\bf 342} (2001) 393.

\bibitem{smp} N.K. Glendenning and J. Schaffner-Bielich, Phys. Rev. 
{\bf C60} (1999) 025803.\\
M. Christiansen and N.K. Glendenning, {\it astro-ph/0008207};
M. Christiansen, N.K. Glendenning and J. Schaffner-Bielich, Phys.
Rev. {\bf C62} (2000) 025804.

\bibitem{pet}
D.G. Ravenhall, C.J. Pethick and J.R. Wilson, Phys. Rev. Lett.
{\bf 50} (1983) 2066;\\
H. Heiselberg, C.J. Pethick and E.F. Staubo, Phys. Rev. Lett. {\bf
70} (1993) 1355.


\bibitem{nor}
T. Norsen and S. Reddy, Phys.Rev. {\bf C63} (2001) 065804.

\bibitem{gug} e.g., E.A. Guggenheim, {\it Thermodynamics}, (North-Holland
        pub., 1977).

\bibitem{vos} D.N. Voskresenky, M. Yasuhira and T. Tatsumi,
Phys. Lett. {\bf B541} (2002) 93, 
; Nucl. Phys. {\bf A723} (2003) 291; T. Tatsumi, M. Yasuhira and
D.N. Voskresenky, Nucl. Phys. {\bf A718} (2003) 359c.

\bibitem{par}
R.G. Parr and W. Yang, {\it Density-Functional Theory of Atoms and
Molecules}, (Oxford U. Press, 1989).

\bibitem{maru} T. Maruyama et al., nucl-th/0311076; in this proceedings.

\end{thebibliography}
\end{document}